\documentclass[twocolumn,fp]{jpsj3}

\usepackage{txfonts}

\usepackage{amsmath}
\usepackage{amsfonts}
\usepackage{bm}
\usepackage{graphicx}
\usepackage{color}

\title{Rossby-Haurwitz wave in a rotating bubble-shaped Bose-Einstein
  condensate}

\author{Hiroki Saito and Masazumi Hayashi}
\inst{Department of Engineering Science, University of
Electro-Communications, Tokyo 182-8585, Japan}

\abst{
A Rossby-Haurwitz (RH) wave is an excitation mode of a fluid on a
rotating spherical surface, which propagates westward in the rotating
frame of reference.
Motivated by the recent realization of the shell-shaped Bose-Einstein
condensate in microgravity, we investigate the RH wave in a superfluid
rotating on a spherical-surface geometry.
We employ the point-vortex model and the three-dimensional
Gross-Pitaevskii equation, and numerically demonstrate that RH waves
can be observed in the system of a superfluid with quantized
vortices.
}

\begin{document}
\date{\today}
\maketitle

\section{Introduction}
\label{s:intro}

In a two-dimensional fluid corotating with a spherical surface, like
the Earth's atmosphere, the vorticity of the fluid depends on the
latitude: the vorticity is largest in the north and south polar
regions while it vanishes near the equator.
This vorticity gradient, combined with the Coriolis effect, produces
intriguing fluid phenomena, such as the dynamics of the meandering jet
stream in the Earth's atmosphere.
In 1939, Rossby and his collaborators identified that such waves under
the vorticity gradient and Coriolis effect can explain several
meteorological observations~\cite{Rossby}, which are nowadays called
Rossby waves~\cite{Platzman, Tritton}.
In 1940, Haurwitz provided the eigenmodes of Rossby waves on a
rotating spherical surface~\cite{Haurwitz}, which are referred to as
Rossby-Haurwitz (RH) waves.

In laboratory experiments, Rossby waves can be realized by water in a
rotating tank with a radially slanted bottom~\cite{Stommel}, which
mimics the vorticity gradient in a rotating sphere.
However, to our knowledge, no experiments involving a fluid on a
rotating spherical surface have been performed in classical systems.
On the other hand, in quantum systems, there have been efforts to
produce a spherical bubble-shaped potential for ultracold
atoms~\cite{Zobay, Zobay2}.
Recently, on the International Space Station under microgravity, a
Bose-Einstein condensate (BEC) confined in a bubble-shaped potential
was produced~\cite{Carollo}.
Such systems have attracted much interest recently, and stimulated
many theoretical~\cite{Padavic17, Sun, Tononi19, Bereta19, Tononi20,
  Padavic20, Moller, Diniz, Kanai, Bereta, Andriati, Rhyno, Arazo,
  Tononi22A, Tononi22, Cara, Wolf} and experimental~\cite{Elliott,
  Lundblad, Guo, Jia} studies.

If we rotate a BEC in a spherical bubble-shaped trap, we may realize a
superfluid analogue of the planetary atmosphere, and various
intriguing phenomena, such as Rossby waves, may be observed. 
Recently, excitation modes in such a rotating superfluid on a
spherical surface were analyzed using a coarse-grained hydrodynamic
model in which quantized vortices were smoothed~\cite{Li}.
Rossby-like waves in a rotating BEC were also studied in
Ref.~\citen{Tercas}.
However, in these studies, quantized vortices were not directly
treated, but were smoothed into continuous fields.

In the present paper, we directly deal with the dynamics of quantized
vortices in a rotating spherical geometry, and numerically demonstrate
RH waves in a superfluid.
We will consider two models: the point-vortex model and the
Gross-Pitaevskii (GP) model.
The point-vortex model is numerically tractable, and we can treat many
quantized vortices with low numerical cost.
We will clearly show that RH waves exist in the point-vortex system
and confirm that the dispersion relation agrees with that in RH
theory.
Supported by these results, we tackle the direct numerical simulation
of the three-dimensional (3D) GP equation for a spherical
bubble-shaped potential, and show that RH waves can be observed in a
BEC in realistic situations.
We will also study the dependence of the dynamics on the interaction
strength and on the thickness of the spherical shell.

This paper is organized as follows.
Section~\ref{s:rossby} reviews RH waves in a classical fluid.
Section~\ref{s:point} studies the point-vortex model on a spherical
surface and identifies the RH waves of point vortices.
Section~\ref{s:GP} shows numerical simulations of the GP equation.
Section~\ref{s:conc} provides conclusions for this study.

\section{Rossby-Haurwitz wave on a rotating sphere}
\label{s:rossby}

In this section, we derive the eigenmode of an RH wave in a
classical fluid~\cite{Rossby, Haurwitz}.
We consider an incompressible inviscid fluid on a 2D sphere with
radius $R$.
We use the polar coordinates $(r, \theta, \phi)$, where $\theta$
represents the usual polar angle, not the latitude, throughout this
paper.
Using the stream function $\Psi(\theta, \phi)$, the flow velocities in
the $\theta$ and $\phi$ directions are expressed as
\begin{equation}
  v_\theta = -\frac{1}{R \sin\theta}
  \frac{\partial\Psi}{\partial\phi},
  \;\;\; v_\phi = \frac{1}{R} \frac{\partial\Psi}{\partial\theta},
\end{equation}
which assures $\nabla \cdot \bm{v} = 0$.
The vorticity $F$ on the spherical 2D surface is written as
\begin{equation}
F = (\nabla \times \bm{v})_r = \nabla^2 \Psi.
\end{equation}
For the static flow that follows the surface of a sphere rotating
around the $z$ axis at a frequency $\Omega$, i.e., for rigid-body
rotation, the flow velocity is given by
\begin{equation} \label{v0}
v_\theta^0 = \Omega R \sin\theta, \;\;\; v_\phi^0 = 0,
\end{equation}
which has the vorticity
\begin{equation} \label{F0}
F^0 = 2\Omega \cos\theta.
\end{equation}

We divide the velocity field into that of the rigid-body rotation and
a small deviation as $\bm{v} = \bm{v}^0 + \delta\bm{v}$.
The vorticity is also divided as $F = F^0 + \nabla^2 \delta\Psi$,
where $\delta\Psi$ is the stream function of $\delta\bm{v}$.
Substituting them into the equation of vorticity conservation,
$\partial F / \partial t + \bm{v} \cdot \nabla F = 0$, we
obtain
\begin{equation} \label{eq0}
\frac{\partial}{\partial t} \nabla^2 \delta\Psi + \bm{v}^0 \cdot
\nabla (\nabla^2 \delta\Psi) + \delta\bm{v} \cdot \nabla F^0 = 0,
\end{equation}
where we use $\bm{v}^0 \cdot \nabla F^0 = 0$ and neglect the second
order term of the small deviation.
Using Eqs.~(\ref{v0}) and (\ref{F0}), Eq.~(\ref{eq0}) becomes
\begin{equation} \label{eq1}
  \left( \frac{\partial}{\partial t} + \Omega
  \frac{\partial}{\partial \phi} \right) \nabla^2 \delta\Psi +
  \frac{2\Omega}{R^2} \frac{\partial\delta\Psi}{\partial\phi} = 0.
\end{equation}
The eigenmode of Eq.~(\ref{eq1}) has the form,
\begin{equation} \label{mode}
\delta\Psi(\theta, \phi, t) \propto Y_\ell^m(\theta, \phi) e^{-i
  \omega t},
\end{equation}
where $Y_\ell^m$ represents the spherical harmonics.
Substituting Eq.~(\ref{mode}) into Eq.~(\ref{eq1}), we obtain the
dispersion relation of the RH wave as
\begin{equation} \label{rossby}
\omega = m \Omega - \frac{2 m \Omega}{\ell (\ell + 1)}.
\end{equation}
Since $Y_\ell^m \propto e^{im\phi}$, the mode in Eq.~(\ref{mode}) is
proportional to $\exp\{i m [\phi - \Omega t +
\frac{2 \Omega t}{\ell (\ell + 1)}]\}$, which indicates that the mode
rotates at the angular frequency
$\Omega - 2\Omega / [\ell (\ell + 1)]$. 
Thus, relative to the rigid-body rotation at $\Omega$, the mode of the
RH wave rotates in the opposite direction (westward for someone
dwelling on the sphere) at the angular frequency~\cite{Haurwitz},
\begin{equation} \label{omegaR}
\frac{2\Omega}{\ell (\ell + 1)} \equiv \omega_R.
\end{equation}

\section{Point-Vortex model}
\label{s:point}
To see RH waves in rotating superfluids with quantized vortices
clearly, we first consider the point-vortex model on a 2D
spherical surface~\cite{Lamb, Bogomolov,Kimura87, Kimura99, Bereta}.
The numerical cost of the point-vortex simulation is much lower than
that of the GP simulation, which enables the treatment of large
numbers of vortices.
In the limit of an infinite density of point vortices, the model
reduces to the classical continuous vorticity discussed in the
previous section.

In the point-vortex model, quantized vortices are represented as
points located at $\bm{r}_j$ on a sphere with radius $R$, where the
subscript $j$ runs from 1 to the total number of point vortices $N_v$.
Each point vortex has a circulation $h q_j / M$, where $M$ is the mass
of an atom and $q_j = +1$ or $-1$,
This approximation is valid for strong confinement of the fluid on a
spherical surface to assure the 2D geometry, and valid for a large
interaction so that the vortex cores become small and the atomic
density at the outside of vortex cores is kept constant.
The numbers of $q_j = +1$ and $-1$ vortices must be the same, from the
topological constraint, and $N_v$ must be even.
The velocity of the fluid at a position $\bm{r}$ on the sphere is
written as
\begin{equation}
  \bm{v}(\bm{r}) = \frac{\hbar}{M R} \bm{r} \times \sum_{j=1}^{N_v}
  q_j \frac{\bm{r} - \bm{r}_j}{|\bm{r} - \bm{r}_j|^2}.
\end{equation}
The point vortex located at $\bm{r}$ moves with the velocity
$\bm{v}(\bm{r})$, and thus the equation of motion of the point
vortices is given by
\begin{equation} \label{eom}
  \dot{\bm{r}}_k = \frac{\hbar}{M R} \bm{r} \times {\sum_{j}}' q_j
  \frac{\bm{r} - \bm{r}_j}{|\bm{r} - \bm{r}_j|^2},
\end{equation}
where $\sum'$ denotes that $j = k$ is excluded from the sum.
Equation~(\ref{eom}) assures that the vortices always stay on the sphere
because $\dot{\bm{r}}_k \perp \bm{r}_k$.

We prepare the initial distribution of the point vortices by the
following procedure.
First, we place the point vortices with vorticity $q_j = \pm1$ at
$\bm{r}_j = (x, y, \pm\sqrt{R^2 - x^2 - y^2})$, where $x$ and $y$ 
obey the uniform random distribution on the $x$-$y$ plane with $x^2 +
y^2 < R^2$, and the $+$ or $-$ sign applies to each half of the $N_v$
vortices.
Hence, the average density of point vortices projected on the $x$-$y$
plane is $N_v / (2 \pi R^2) \equiv n_v$ for each of the northern and
southern hemispheres, and the average number of vortices contained in
a circle of a latitude line at $\theta$ is $\pi (R \sin\theta)^2 n_v
\equiv N_\theta$.
Thus, on average, the flow velocity in the direction of a latitude
line is given by $\hbar / M \cdot 2\pi N_\theta  / (2\pi R
\sin\theta) = \hbar N_v \sin\theta / (2MR)$, which corresponds to
rigid-body rotation of the sphere about the $z$ axis at an angular
frequency
\begin{equation} \label{Omega}
  \Omega = \frac{\hbar N_v}{2 M R^2}.
\end{equation}

The random distribution of point vortices contains an extra excitation
energy arising from the randomness, which obscures the RH waves, and
we must lower the energy.
The energy of the point-vortex system is written as~\cite{Bereta}
\begin{equation}
E_{\rm pv} = E_{\rm self} - \frac{\pi \hbar^2 n}{M} \sum_{j \neq k}
q_j q_k \log \frac{|\bm{r}_j - \bm{r}_k|}{R},
\end{equation}
where $E_{\rm self}$ is the self-energy of the vortices, which is
independent of their positions, and $n$ is the atomic density.
The effective energy in the rotating frame is defined by $E_{\rm pv} -
\Omega L_z \equiv K$, where
\begin{equation}
  L_z = M n \int [\bm{r} \times \bm{v}(\bm{r})]_z d^2\bm{r}
  = 2 \pi n \hbar R \sum_j q_j z_j
\end{equation}
is the angular momentum in the $z$ direction.
We minimize $K$ by evolving the vortex positions $\bm{r}_k$ as
\begin{eqnarray} \label{iter}
\bm{r}_k & \rightarrow & P\left[ \bm{r}_k - \epsilon \frac{\partial
    K}{\partial\bm{r}_k} \right] \nonumber \\
& = & P\left[ \bm{r}_k + \epsilon \frac{2\pi \hbar^2 n}{M} {\sum_{j}}'
  q_k q_j \frac{\bm{r}_k - \bm{r}_j}{|\bm{r}_k - \bm{r}_j|^2} + 2\pi n
  \hbar R q_k \hat{z} \right],
\nonumber \\
\end{eqnarray}
where $\epsilon > 0$ is a small number, $P[\cdot]$ represents the
projection of the vector $\bm{r}_k + \Delta\bm{r}_k$ on the spherical
surface, and $\hat{z}$ is the unit vector in the $z$ direction.
By sufficient iterations of Eq.~(\ref{iter}), we obtain the stationary
distribution of point vortices in a frame rotating at $\Omega$.
We have confirmed numerically that the obtained state is stationary under
Eq.~(\ref{eom}) in the rotating frame.

We excite the RH wave on the stationary state obtained above.
We give a small shift to the azimuthal angle of each position
$\bm{r}_k$ as
\begin{equation} \label{shift}
\delta\phi_k = c \sin^m\theta_k \sin(m\phi_k),
\end{equation}
where $\theta_k$ and $\phi_k$ are the polar and azimuthal angles of
$\bm{r}_k$, and $c$ is a constant.
Since the average vortex density before the shift is $\propto 2 \Omega
\cos\theta$, the change in the average vortex density caused by the
shift $\delta\phi_k$ is $\propto -\sin^m\theta \cos\theta \cos(m\phi)
\propto {\rm Re}[Y_{m+1}^m(\theta, \phi)]$.
Thus, we can excite the eigenmode of the spherical RH wave in
Eq.~(\ref{mode}) with $\ell = m + 1$.

\begin{figure}[tb]
\includegraphics[width=8.5cm]{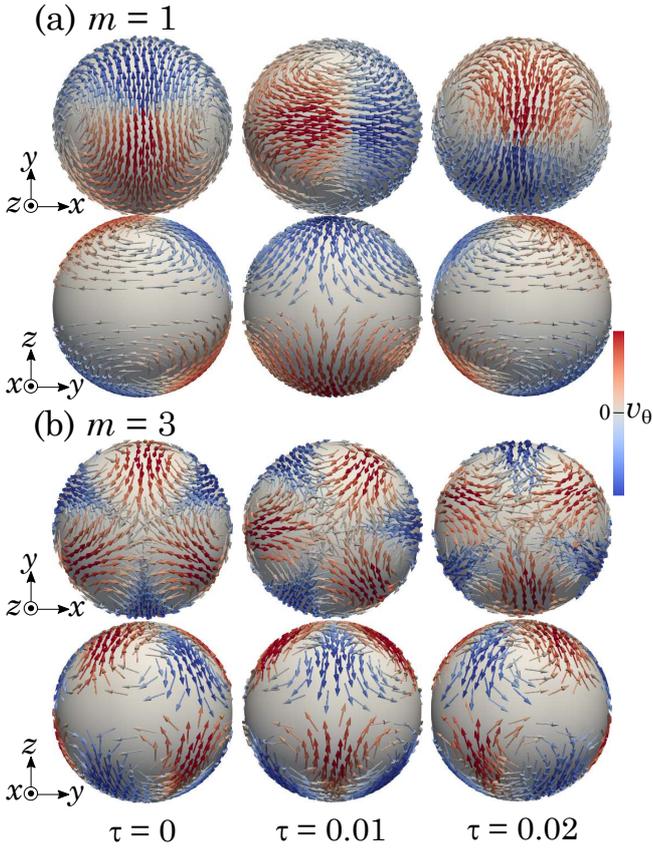}
\caption{
  Time evolution of the point vortices with $N_v = 1000$ in the
  rotating frame of reference at frequency $\Omega = \hbar N_v /
  (2MR^2)$.
  The initial perturbation is given as Eq.~(\ref{shift}) for (a) $m =
  1$ and (b) $m = 3$ with $c = 0.1$.
  The upper and lower rows show the distributions seen from the $+z$
  and $+x$ directions, respectively.
  The position, direction, and color of each arrow indicate
  $\bm{r}_j$, $\dot{\bm{r}}_j$, and $v_\theta = \dot{\bm{r}}_j \cdot
  \hat{\bm{e}}_\theta$ of each point vortex, respectively, where
  $\hat{\bm{e}}_\theta$ is the unit vector in the meridian direction
  at $\bm{r}_j$.
  Time is normalized as $\tau = \hbar t / (MR^2)$.
  See the Supplemental Material for videos of the
  dynamics~\cite{movies}.
}
\label{f:pv_ev}
\end{figure}
Starting from this initial distribution of the point vortices, we
solve the equation of motion in Eq.~(\ref{eom}) using the fourth-order
Runge-Kutta method.
Figure~\ref{f:pv_ev} shows the time evolution of $N_v = 1000$ vortices
with the initial perturbation in Eq.~(\ref{shift}) with $m = 1$ and $m
= 3$.
The vortex density is proportional to $\cos\theta$ and the vortices
are sparse near the equator, as seen in Fig.~\ref{f:pv_ev}.
For $m = 1$ in Fig.~\ref{f:pv_ev}(a), the initial perturbation in
Eq.~(\ref{shift}) decreases (increases) the vortex density around
$\phi = 0$ ($\phi = \pi$).
As a result, clockwise (counterclockwise) flow is induced around $\phi
= 0$ ($\phi = \pi$) on the northern ($z > 0$) hemisphere, and an
oppositely circulating flow is induced on the southern hemisphere.
As time evolves, one can see that this flow pattern rotates around the
$z$ axis.
Since Fig.~\ref{f:pv_ev} is shown in the frame rotating at $\Omega$,
this evolution of the flow pattern agrees with the dispersion relation
in Eq.~(\ref{omega}), i.e., the flow pattern moves westward for those
dwelling on the spherical surface.
Also, for $m = 3$ in Fig.~\ref{f:pv_ev}(b), where there are six
circulating flows on each of the northern and southern hemispheres,
the flow pattern moves westward in a similar manner.
It should be noted that the westward movement of the flow pattern is
not caused by the westward movement of the point vortices themselves.
This can be clearly seen in the videos in the Supplemental
Material~\cite{movies}, in which the positions of the point vortices
are not much changed compared with the movement of the patterns.

\begin{figure}[tb]
\includegraphics[width=8cm]{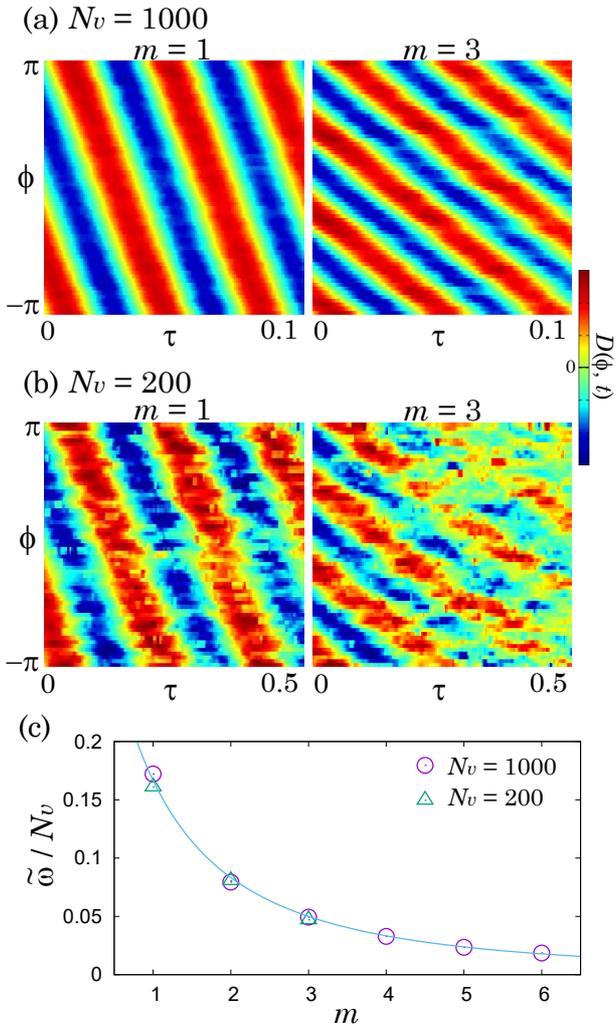}
\caption{
Time evolution of the azimuthal distribution $D(\phi, t)$ of vortex
flow in the meridian direction defined in Eq.~(\ref{D}) for (a) $N_v =
1000$ and (b) $N_v = 200$.
Time is normalized as $\tau = \hbar t / (MR^2)$.
(c) Normalized rotation frequency $\tilde\omega = M R^2 \omega /
\hbar$ of the RH wave as a function of $m$ for $N_v = 1000$ (circles)
and $N_v = 200$ (triangles).
The solid line indicates Eq.~(\ref{omega}).
}
\label{f:pv_omega}
\end{figure}
The rotation of the flow patterns can be visualized more clearly by the
azimuthal distribution of the averaged meridian flow defined by
\begin{equation} \label{D}
  D(\phi, t) = \frac{1}{N_\phi} \sum_{[\phi, \phi+\Delta\phi]} 
  q_j \dot{\bm{r}}_j(t) \cdot \hat{e}_\theta,
\end{equation}
where the summation is taken over the point vortices inside the narrow
azimuthal region from $\phi$ to $\phi + \Delta\phi$, $N_\phi$ is the
number of vortices in this region, and $\hat{e}_\theta$ is the unit
vector in the meridian direction at $\bm{r}_j$.
The vorticity $q_j$ is needed in the summand, otherwise contributions
from the northern and southern hemispheres cancel (see the flow
patterns in Fig.~\ref{f:pv_ev}).
In Eq.~(\ref{D}), all the quantities are taken in the frame rotating
at $\Omega$.
Figures~\ref{f:pv_omega}(a) and \ref{f:pv_omega}(b) show the time
evolution of $D(\phi, t)$ for $N_v = 1000$ and $N_v = 200$.
For $N_v = 1000$ in Fig.~\ref{f:pv_omega}(a), the westward movement of
the flow pattern is clearly seen both for $m = 1$ and $m = 3$.
However, for $N_v = 200$ in Fig.~\ref{f:pv_omega}(b), the distribution
is noisy because of the small number of point vortices, and the
pattern rapidly decays for larger $m$.

From the slopes of the strip-shaped distributions in
Figs.~\ref{f:pv_omega}(a) and \ref{f:pv_omega}(b), we obtain the
angular frequency of the westward movement of the flow pattern, which
is plotted in Fig.~\ref{f:pv_omega}(c).
From Eqs.~(\ref{omegaR}) and (\ref{Omega}), the RH-wave frequency is
written as
\begin{equation} \label{omega}
\omega_R = \frac{2\Omega}{\ell (\ell + 1)} = \frac{\hbar N_v}{M R^2
  (m + 1) (m + 2)}
\end{equation}
under the excitation of the $Y_{m+1}^m$ mode.
Equation~(\ref{omega}) is shown by the solid line in
Fig.~\ref{f:pv_omega}(c), which agrees well with the plots.
This indicates that RH waves are successfully excited in the dynamics
of the point-vortex system, shown in Figs.~\ref{f:pv_ev} and
\ref{f:pv_omega}.
The results also imply that an RH wave can be observed in the system
of quantized vortices, when the number of quantized vortices is
$\gtrsim 100$.
However, for $N_v \sim 100$, the RH wave is unclear for large $m$,
which restricts the plots in Fig.~\ref{f:pv_omega}(c) to $m \leq 3$.

\section{Numerical simulations of the Gross-Pitaevskii equation}
\label{s:GP}

To investigate a more realistic system of a BEC on a spherical
bubble-shaped geometry, we perform numerical simulations of the 3D GP
equation.
In the mean-field approximation, a BEC at zero temperature can be
described by the macroscopic wave function $\psi(\bm{r}, t)$, where
the normalization condition is given by $\int |\psi|^2 d\bm{r} = N$
with $N$ being the number of atoms.
In the rotating frame of reference, the macroscopic wave function
obeys the GP equation as
\begin{equation} \label{GP}
  i \hbar \frac{\partial \psi}{\partial t} = -\frac{\hbar^2}{2M}
  \nabla^2 \psi - \Omega \hat{L}_z \psi + V \psi
  + \frac{4\pi\hbar^2 a}{M} |\psi|^2 \psi,
\end{equation}
where $\Omega$ is the angular frequency of rotation, $\hat{L}_z = -i
\hbar (x \partial / \partial y - y \partial / \partial x)$ is the $z$
component of the angular momentum operator, and $a$ is the $s$-wave
scattering length of atoms.
The external potential $V$ consists of three parts: $V(\bm{r}, t) =
V_s(\bm{r}) + V_c(\bm{r}) + V_p(\bm{r}, t)$.
The spherical bubble-shaped trap potential is approximated to be a
harmonic potential around the radius $R$ as
\begin{equation}
  V_s(\bm{r}) = \frac{1}{2} M \omega_t^2 (\sqrt{x^2 + y^2 + z^2} -
  R)^2,
\end{equation}
where $\omega_t$ is the radial trap frequency.
We also add a potential expressed as
\begin{equation} \label{Vc}
  V_c(\bm{r}) = \frac{1}{2} M \Omega^2 (x^2 + y^2),
\end{equation}
which compensates the centrifugal force and prevents the atomic cloud
from concentrating around the equator.
We add a perturbation potential to excite the RH wave as
\begin{equation} \label{Vp}
  V_p(\bm{r}, t) = \alpha(t) (x^2 + y^2)^{m/2} \cos(m \phi),
\end{equation}
where $\alpha(t)$ is a function of time specified later and $m$ is a
positive integer.
The perturbation potential $V_p$ induces inhomogeneity in the density,
which affects the vortex distribution.
Since the vortex density on the rotating spherical surface is $\propto
\cos\theta$ and $(x^2 + y^2)^{m/2} \propto \sin^m \theta$, we expect
that the potential in Eq.~(\ref{Vp}) can excite the RH mode in
Eq.~(\ref{mode}) with ${\rm Re} [Y_{m+1}^m] \propto \sin^m \theta \cos
\theta \cos(m\phi)$.
The potentials in the forms of Eqs.~(\ref{Vc}) and (\ref{Vp}) can be
produced by Gaussian and Laguerre-Gaussian laser beams propagating in
the $z$ direction.

\begin{figure}[tb]
\includegraphics[width=7.5cm]{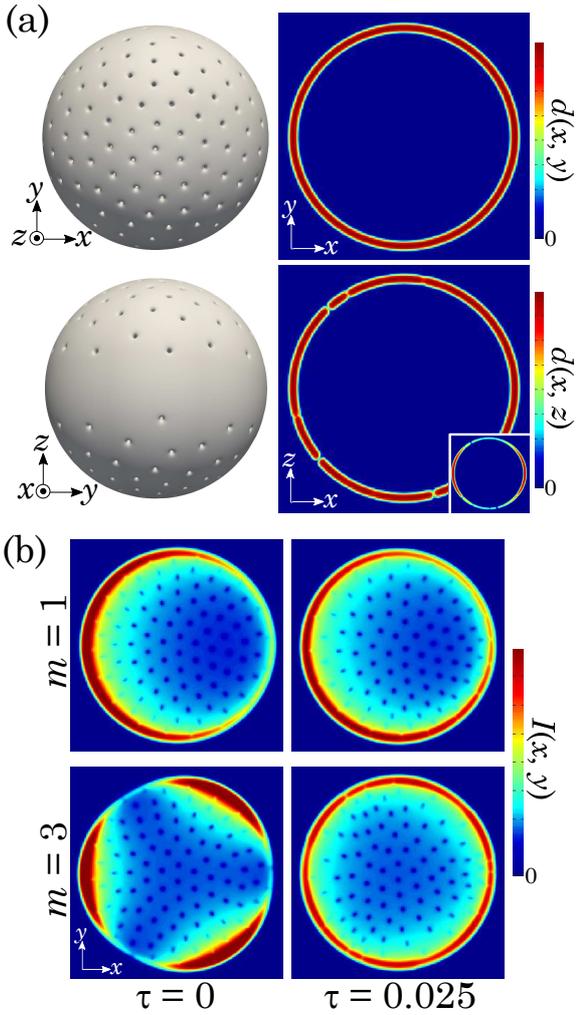}
\caption{
  (a) Stationary state obtained by the imaginary-time evolution of the
  GP equation without the perturbation potential, $V_p = 0$.
  The left-hand spheres show the isodensity surfaces at half the
  maximum density, as seen from the $+z$ and $+x$ directions.
  The right-hand panels show the density distributions on the $z = 0$
  and $y = 0$ planes, $d(x, y) = |\psi(x, y, 0)|^2$ and $d(x, z) =
  |\psi(x, 0, z)|^2$.
  The inset shows the case without $V_c$.
  (b) Early stage of the real-time evolution.
  The integrated density $I(x, y) = \int |\psi|^2 dz$ is shown.
  The initial state is prepared with the perturbation potential in
  Eq.~(\ref{Vp}), where $\tilde\alpha \equiv \alpha M R^{m+2} /
  \hbar^2 = 2500$ for $m = 1$ and $\tilde\alpha = 4000$ for $m = 3$.
  In the real-time evolution, the perturbation potential is ramped
  down as in Eq.~(\ref{ramp}) with $t_{\rm off} = 0.025 M R^2 /
  \hbar$.
  Time is normalized as $\tau = t \hbar / (M R^2)$.
  In (a) and (b), the parameters are $\tilde\Omega = M R^2 \Omega /
  \hbar = 100$, $\tilde\omega_t = M R^2 \omega_t / \hbar = 2000$, and
  $4\pi a N / R = 7000$.
}
\label{f:GPstat}
\end{figure}
We solve the 3D GP equation~(\ref{GP}) in Cartesian coordinates
using the pseudospectral method~\cite{recipe} with a fourth-order
Runge-Kutta scheme.
We discretize space with size $(2.3R)^3$ into a $256^3$ mesh and the
typical time step is $dt = 10^{-6}$.
The stationary state in the rotating frame is obtained by the
imaginary-time propagation of the GP equation, where $i$ on the
left-hand side is replaced by $-1$.
Figure~\ref{f:GPstat}(a) shows the stationary state for $\tilde\Omega
= M R^2 \Omega / \hbar = 100$ without the perturbation potential, $V_p
= 0$.
The density distribution is well localized near the radius $R$ and the
system is approximately two-dimensional.
The BEC contains quantized vortices on the sphere; the vorticity is
positive (negative) on the northern (southern) hemisphere and the
vortices are sparse near the equator, in a manner similar to the case
of the point-vortex model (see Fig.~\ref{f:pv_ev}).
The total number of vortices is $99 \times 2$, which is in good
agreement with $N_v = 2 M R^2 \Omega / \hbar = 200$ in
Eq.~(\ref{Omega}).
One can see that the vortex distribution is not a perfect triangular
lattice, which is due to the curvature of the surface.
If the compensation potential $V_c$ in Eq.~(\ref{Vc}) is absent, the
atoms accumulate in the equatorial region, as shown in the inset in
Fig.~\ref{f:GPstat}(a).

To observe the RH wave, we include the perturbation potential $V_p$ in
Eq.~(\ref{Vp}) in the imaginary-time evolution.
The resulting stationary states are shown in Fig.~\ref{f:GPstat}(b)
(panels at $\tau = 0$), where the density distributions exhibit
$m$-fold modulation.
Figure~\ref{f:GPstat}(b) shows the early stage of the real-time
evolution of the initially modulated states.
In the real-time evolution, the perturbation potential $V_p$ is
linearly switched off as
\begin{equation} \label{ramp}
  \alpha(t) = \left\{ \begin{array}{ll}
    \alpha(0) (1 - t / t_{\rm off}) & (t < t_{\rm off}) \\
    0 & (t \geq t_{\rm off}), \end{array} \right.
\end{equation}
where $t_{\rm off}$ is the ramp-down time.
This gradual elimination of $V_p$ reduces extra excitation of the
system.
It can be seen from Fig.~\ref{f:GPstat}(b) (at $\tau = 0.025$) that
the vortex distributions exhibit $m$-fold modulation.
For $t \geq t_{\rm off}$, the system evolves in the axisymmetric
potential.

\begin{figure}[tb]
\includegraphics[width=8cm]{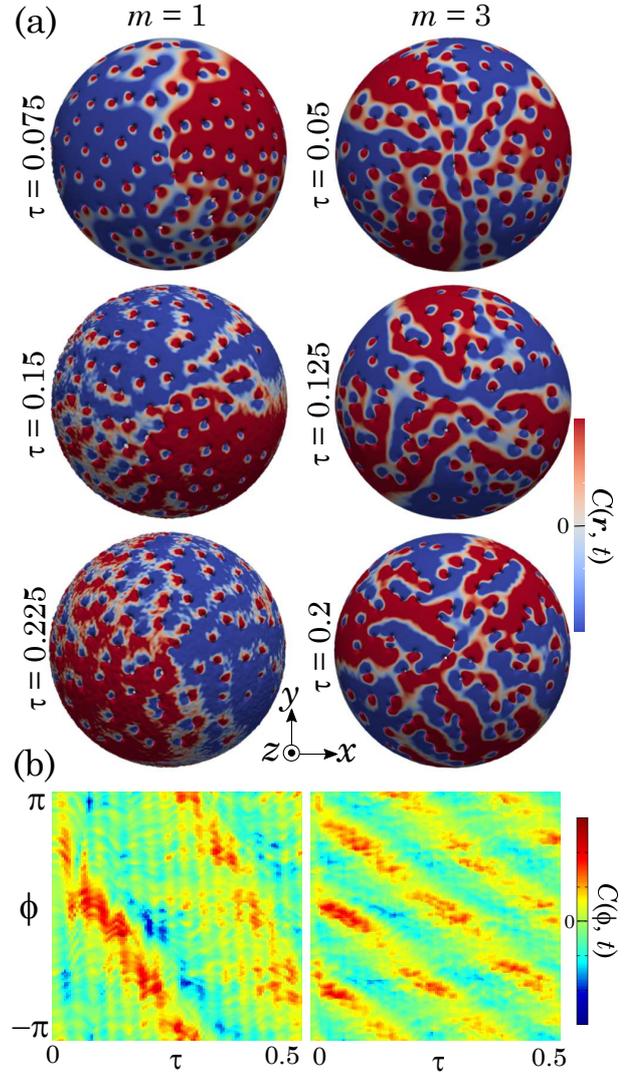}
\caption{
  (a) Time evolution following Fig.~\ref{f:GPstat}(b).
  The isodensity surface is colored by the mass current in the
  meridian direction $C(\bm{r}, t)$ defined in Eq.~(\ref{F}).
  (b) Azimuthal distributions of the meridian flow $C(\phi, t)$
  defined in Eq.~(\ref{Cphi}).
  All the results are shown in the frame rotating at the frequency
  $\Omega$.
  See the Supplemental Material for videos of the
  dynamics~\cite{movies}.
}
\label{f:GPev}
\end{figure}
Figure~\ref{f:GPev} shows the real-time evolution, which follows
Fig.~\ref{f:GPstat}(b).
The color on the isodensity surface represents the mass current in the
meridian direction,
\begin{equation} \label{F}
  C(\bm{r}, t) = \frac{\hbar}{M} {\rm Im} \left[ \psi^*(\bm{r}, t)
  \nabla \psi(\bm{r}, t) \right] \cdot \hat{e}_\theta.
\end{equation}
Apart from the rapid circulation (steep change in the sign of
$C(\bm{r}, t)$) near the quantized vortices, the global flow
patterns in Fig.~\ref{f:GPev}(a) exhibit $m$-fold modulation, and
the flow patterns rotate westward in the frame rotating at the
frequency $\Omega$ both for $m = 1$ and $m = 3$.
The westward movement of the flow pattern is not caused by that of the
vortices (see the videos in the Supplemental Material), in a manner
similar to the point-vortex model in Fig.~\ref{f:pv_ev}.
We define the azimuthal distribution of the meridian flow as
\begin{equation} \label{Cphi}
  C(\phi, t) = \int C(\bm{r}, t) {\rm sign}(z) r^2 \sin\theta dr
  d\theta,
\end{equation}
where the sign of $z$ is multiplied since the meridian flows in the
northern and southern hemispheres are opposite to each other for the
modes excited in Fig.~\ref{f:GPev}.
Figure~\ref{f:GPev}(b) shows $C(\phi, t)$, which clearly indicates the
westward movements of the flow patterns.
The angular frequencies $\omega_R$ of the westward rotation can be
obtained from Fig.~\ref{f:GPev}(b), which are $\tilde{\omega}_R \equiv
M R^2 \omega_R / \hbar \simeq 20$ for $m = 1$ and $\tilde{\omega}_R
\simeq 9$ for $m = 3$.
The RH-wave frequencies predicted in Eq.~(\ref{omegaR}) with $\ell = m
+ 1$ are $\tilde{\omega}_R = 100 / 3$ for $m = 1$ and
$\tilde{\omega}_R = 10$ for $m = 3$.
Thus, the GP result agrees well with the RH theory for $m = 3$,
whereas there is a discrepancy between them for $m = 1$.

\begin{figure}[tb]
\includegraphics[width=8cm]{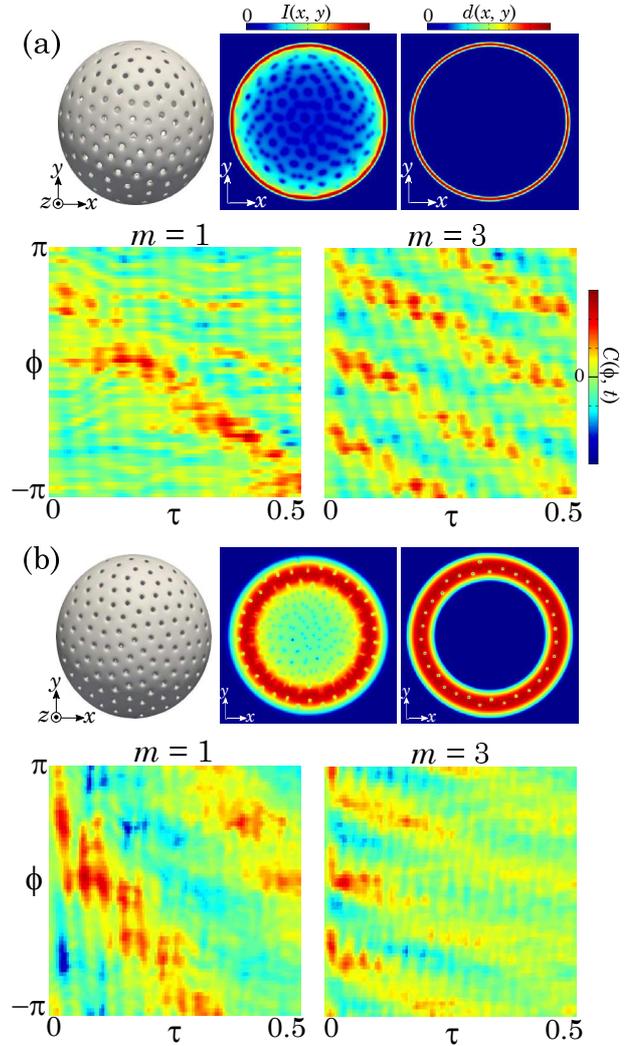}
  \caption{
    (a) $4\pi a N / R$ and $\tilde\alpha$ are ten times smaller than
    those in Fig.~\ref{f:GPstat}.
    (b) $\tilde\omega_t$ is sixteen time smaller and $4\pi a N / R$ is
    four times larger than those in Fig.~\ref{f:GPstat}.
  Top rows (from left to right): Integrated density $I(x, y) = \int
  |\psi|^2 dz$, cross-sectional density on the $z = 0$ plane $d(x, y)
  = |\psi(x, y, 0)|^2$, and isodensity surface at half the maximum
  density of the stationary state without the perturbation potential,
  $V_p = 0$.
  Bottom rows: Azimuthal distributions of the meridian flow $C(\phi,
  t)$ for the initial perturbation with $m = 1$ and $m = 3$.
  The perturbation potential is linearly ramped down as in
  Fig.~\ref{f:GPstat}.
}
\label{f:param}
\end{figure}
To understand the origin of the above discrepancy, we study the
parameter dependence.
When the vortex core becomes larger, the density distribution becomes
more inhomogeneous, and the assumption made in deriving the classical
RH wave breaks down.
We examine this effect by decreasing the interaction strength.
Figure~\ref{f:param}(a) shows the case of an interaction strength ten
times smaller than that in Figs.~\ref{f:GPstat} and \ref{f:GPev}.
One can see that the vortex cores are larger and the width of the
spherical shell is thinner than those in Fig.~\ref{f:GPstat}.
It can be found from the dynamics of $C(\phi, t)$ in
Fig.~\ref{f:param}(a) that $\tilde\omega_R \simeq 10$ for $m = 1$ and
$\tilde\omega_R \simeq 8$ for $m = 3$;
the westward movement of the RH wave is retarded particularly for $m =
1$.

When the thickness of the shell is comparable to the radius $R$, the
2D assumption breaks down.
Figure~\ref{f:param}(b) shows the case with a thicker spherical shell
(about four times thicker than Fig.~\ref{f:GPstat}) using a smaller
radial trap frequency $\omega_t$.
Even for this shell thickness, the RH waves still survive, although
they decay faster than in Fig.~\ref{f:GPev}.
The RH-wave frequencies in Fig.~\ref{f:param}(b) are $\tilde\omega_R
\simeq 16$ for $m = 1$ and $\tilde\omega_R \simeq 7$ for $m = 3$, which
are slightly smaller than those in Fig.~\ref{f:GPev}.
The maximum density and then the minimum healing length in
Fig.~\ref{f:param}(b) is almost the same as that in
Fig.~\ref{f:GPstat}(a).
Thus, the RH-wave propagation is made slower for larger vortex cores
and shell thickness, whose mechanism should be further investigated.

Finally, we discuss experimental situations.
Assuming that the $s$-wave scattering length is 100 Bohr radii and the
bubble radius $R = 10$ $\mu{\rm m}$, the number of atoms needed is $N
\simeq 1.05 \times 10^6$ for the condition in Figs.~\ref{f:GPstat} and
\ref{f:GPev}, and $N \simeq 1.05 \times 10^5$ for Fig.~\ref{f:param},
which is possible in current experimental setups.
Using the mass of a $^{87}{\rm Rb}$ atom, we can estimate the rotation
and radial trap frequencies used in Figs.~\ref{f:GPstat} and
\ref{f:GPev} as $\Omega \simeq 2\pi \times 117$ Hz and $\omega_t
\simeq 2\pi \times 2.3$ kHz, and the normalized time $\tau = 0.5$
corresponds to 68 ms.
The radius $R$ and the radial trap frequency $\omega_t$ can be
controlled by choosing the strengths of the DC and AC magnetic fields
producing the bubble-shaped trap~\cite{Zobay, Zobay2}.
In experiments, it is difficult to measure the mass-current
distribution.
If nondestructive measurement of vortex positions is possible, the
flow velocity can be obtained from the movement of the vortices.
The real-time imaging technique of vortices may be used for this
purpose~\cite{Freilich, Middel}.

\section{Conclusions}
\label{s:conc}

We have investigated RH waves in a superfluid rotating on a spherical
surface using the point-vortex model and the Gross-Pitaevskii
equation.
In Sec.~\ref{s:point}, we employed the point-vortex model, and
numerically solved the equation of motion for point vortices on a 2D
spherical surface.
We demonstrated that the eigenmodes of the RH waves can be excited in
the quantized vortex system, and showed that the RH waves persist for
a long time when the number of vortices $N_v$ is large and the order
of the eigenmode $\ell = m + 1$ is small.
The frequency of the westward rotation of the RH wave agreed well with
the analytical prediction.
In Sec.~\ref{s:GP}, we numerically solved the full-3D GP equation for
a spherical bubble-shaped potential, and observed RH waves excited in
this system.
We confirmed that the RH-wave frequency in the GP model is in good
agreement with the analytical prediction in classical fluids.
We also studied the dependence of the RH-wave frequency on the
interaction strength and the thickness of the spherical shell.

Direct GP simulations will be helpful for studying other planetary
waves, such as Kelvin and Yanai waves, in superfluids, and may confirm
the topological properties predicted in Ref.~\citen{Li}, which will
be an interesting research direction in the future.

\begin{acknowledgments}
We wish to thank A. Maruyama for contributing to the early stage of this
work.
This work was supported by JSPS KAKENHI Grant Number JP20K03804.
\end{acknowledgments}


\begin{thebibliography}{99}

\bibitem{Rossby}
  C. -G. Rossby {\it et al.},
  Relation between variations in the intensity of the zonal
  circulation of the atmosphere and the displacements of the
  semi-permanent centers of action,
  J. Mar. Res. {\bf 2}, 38 (1939).

\bibitem{Platzman}
  G. W. Platzman,
  The Rossby wave,
  Quart. J. Roy. Met. Soc. {\bf 94}, 225 (1968).

\bibitem{Tritton}
  D. J. Tritton,
  {\it Physical Fluid Dynamics}, 2nd ed.
  (Oxford Science Publications, Oxford, 1988).

\bibitem{Haurwitz}
  B. Haurwitz,
  The motion of atmospheric disturbances on the spherical earth,
  J. Mar. Res. {\bf 3}, 254 (1940).

\bibitem{Stommel}
  H. Stommel, A. B. Arons, and A. J. Faller,
  Some examples of stationary planetary flow patterns in bounded
  basins,
  Tellus {\bf 10}, 179 (1957).

\bibitem{Zobay}
  O. Zobay and B. M. Garraway,
  Two-dimensional atom trapping in field-induced adiabatic potentials,
  Phys. Rev. Lett. {\bf 86}, 1195 (2001).

\bibitem{Zobay2}
  O. Zobay and B. M. Garraway,
  Atom trapping and two-dimensional Bose-Einstein condensates in
  field-induced adiabatic potentials,
  Phys. Rev. a {\bf 69}, 023605 (2004).

\bibitem{Carollo}
  R. A. Carollo, D. C. Aveline, B. Rhyno, S. Vishveshwara, C. Lannert,
  J. D. Murphree, E. R. Elliott, J. R. Williams, R. J. Thompson, and
  N. Lundblad,
  Observation of ultracold atomic bubbles in orbital microgravity,
  Nature (London) {\bf 606}, 281 (2022).

\bibitem{Padavic17}
  K. Padavi\'{c}, K. Sun, C. Lannert, and S. Vishveshwara,
  Physics of hollow Bose-Einstein condensates,
  EPL {\bf 120}, 20004 (2017).

\bibitem{Sun}
  K. Sun, K. Padavi\'{c}, F. Yang, S. Vishveshwara, and C. Lannert,
  Static and dynamic properties of shell-shaped condensates,
  Phys. Rev. A {\bf 98}, 013609 (2018).

\bibitem{Tononi19}
  A. Tononi and L. Salasnich,
  Bose-Einstein condensation on the surface of a sphere,
  Phys. Rev. Lett. {\bf 123}, 160403 (2019).

\bibitem{Bereta19}
  S. J. Bereta, L. Madeira, V. S. Bagnato, and M. A. Caracanhas,
  Bose-Einstein condensation in spherically symmetric traps,
  Am. J. Phys. {\bf 87}, 924 (2019).

\bibitem{Tononi20}
  A. Tononi, F. Cinti, and L. Salasnich,
  Quantum bubbles in microgravity,
  Phys. Rev. Lett. {\bf 125}, 010402 (2020).

\bibitem{Padavic20}
  K. Padavi\'{c}, K. Sun, C. Lannert, and S. Vishveshwara,
  Vortex-antivortex physics in shell-shaped Bose-Einstein condensates,
  Phys. Rev. A {\bf 102}, 043305 (2020).
  
\bibitem{Moller}
  N. S. M\'{o}ller, F. E. A. dos Santos, V. S. Bagnato, and
  A. Pelster,
  Bose-Einstein condensation on curved manifolds,
  New J. Phys. {\bf 22}, 063059 (2020).

\bibitem{Diniz}
  P. C. Diniz, E. A. B. Oliveira, A. R. P. Lima, and  E. A. L. Henn,
  Ground state and collective excitations of a dipolar Bose-Einstein
  condensate in a bubble trap,
  Sci. Rep. {\bf 10}, 4831 (2020).

\bibitem{Kanai}
  T. Kanai and W. Guo,
  True mechanism of spontaneous order from turbulence in
  two-dimensional superfluid manifolds,
  Phys. Rev. Lett. {\bf 127}, 095301 (2021).

\bibitem{Bereta}
  S. J. Bereta, M. A. Caracanhas, and A. L. Fetter,
  Superfluid vortex dynamics on a spherical film,
  Phys. Rev. A {\bf 103}, 053306 (2021).

\bibitem{Andriati}
  A. Andriati, L. Brito, L. Tomio, and A. Gammal,
  Stability of a Bose-condensed mixture on a bubble trap,
  Phys. Rev. A {\bf 104}, 033318 (2021).

\bibitem{Rhyno}
  B. Rhyno, N. Lundblad, D. C. Aveline, C. Lannert, and
  S. Vishveshwara,
  Thermodynamics in expanding shell-shaped Bose-Einstein condensates,
  Phys. Rev. A {\bf 104}, 063310 (2021).

\bibitem{Arazo}
  M. Arazo, R. Mayol, and M. Guilleumas,
  Shell-shaped condensates with gravitational sag: contact and dipolar
  interactions,
  New J. Phys. {\bf 23}, 113040 (2021).

\bibitem{Tononi22A}
  A. Tononi,
  Scattering theory and equation of state of a spherical
  two-dimensional Bose gas,
  Phys. Rev. A {\bf 105}, 023324 (2022).

\bibitem{Tononi22}
  A. Tononi, A. Pelster, and L. Salasnich,
  Topological superfluid transition in bubble-trapped condensates,
  Phys. Rev. Res. {\bf 4}, 013122 (2022).

\bibitem{Cara}
  M. A. Caracanhas, P. Massignan, and A. L. Fetter,
  Superfluid vortex dynamics on an ellipsoid and other surfaces of
  revolution,
  Phys. Rev. A {\bf 105}, 023307 (2022).

\bibitem{Wolf}
  A. Wolf, P. Boegel, M. Meister, A. Bala\v{z}, N. Gaaloul, and M. A. Efremov,
  Shell-shaped Bose-Einstein condensates realized with dual-species mixtures,
  Phys. Rev. A {\bf 106}, 013309 (2022).

\bibitem{Elliott}
  E. R. Elliott, M. C. Krutzik, J. R. Williams, R. J. Thompson, and
  D. C. Aveline,
  NASA's Cold Atom Lab (CAL): system development and ground test
  status,
  npj Microgravity {\bf 4}, 16 (2018).

\bibitem{Lundblad}
  N. Lundblad, R. A. Carollo, C. Lannert, M. J. Gold, X. Jiang,
  D. Paseltiner, N. Sergay, and D. C. Aveline,
  Shell potentials for microgravity Bose-Einstein condensates,
  npj Microgravity {\bf 5}, 30 (2019).

\bibitem{Guo}
  Y. Guo, E. M. Gutierrez, D. Rey, T. Badr, A. Perrin, L. Longchambon,
  V. S. Bagnato, H. Perrin, and R. Dubessy,
  Expansion of a quantum gas in a shell trap,
  New J. Phys. {\bf 24}, 093040 (2022).

\bibitem{Jia}
  F. Jia, Z. Huang, L. Qiu, R. Zhou, Y. Yan, and D. Wang,
  Expansion dynamics of a shell-shaped Bose-Einstein condensate,
  Phys. Rev. Lett. {\bf 129}, 243402 (2002).

\bibitem{Li}
  G. Li and D. K. Efimkin,
  Equatorial waves in rotating bubble-trapped superfluids,
  arXiv:2210.10525 (2022).

\bibitem{Tercas}
  H. Ter\c{c}as, J. P. A. Martins, and J. T. Mendon\c{c}a,
  Shell potentials for microgravity Bose-Einstein condensates,
  New J. Phys. {\bf 12}, 093001 (2010).

\bibitem{Lamb}
  H. Lamb, {\it Hydrodynamics}, 6th ed.
  (Dover Publications, New York, 1945).
  
\bibitem{Bogomolov}
  V. A. Bogomolov,
  Dynamics of vorticity at a sphere,
  Fluid Dyn. {\bf 12}, 863 (1977).

\bibitem{Kimura87}
  Y. Kimura and H. Okamoto,
  Vortex motion on a sphere,
  J. Phys. Soc. Jpn. {\bf 56}, 4203 (1987).

\bibitem{Kimura99}
  Y. Kimura and H. Okamoto,
  Vortex motion on surfaces with constant curvature,
  Proc. R. Soc. Lond. A {\bf 455}, 245 (1999).

\bibitem{movies}
See Supplemental Material for videos of the dynamics is provided
online.

\bibitem{recipe}
W. H. Press, S. A. Teukolsky, W. T. Vetterling, and B. P. Flannery,
{\it Numerical Recieps}, 3rd ed. (Cambridge Univ. Press, Cambridge, 2007).  

\bibitem{Freilich}
  D. V. Freilich, D. M. Bianchi, A. M. Kaufman, T. K. Langin, and
  D. S. Hall,
  Real-time dynamics of single vortex lines and vortex dipoles in a
  Bose-Einstein condensate,
  Science {\bf 329}, 1182 (2010).

\bibitem{Middel}
  S. Middelkamp, P. J. Torres, P. G. Kevrekidis, D. J. Frantzeskakis,
  R. Carretero-Gonz\'{a}lez, P. Schmelcher, D. V. Freilich, and
  D. S. Hall,
  Guiding-center dynamics of vortex dipoles in Bose-Einstein
  condensates,
  Phys. Rev. A {\bf 84}, 011605(R) (2011).


\end{thebibliography}
\end{document}